\documentclass[12pt,epsf]{iopart}

\usepackage{pstricks}
\usepackage{pst-coil}
\usepackage{amssymb}
\usepackage{graphics,epsf}

\def\p{\partial}
\def\sup{\Sigma}
\def\suup{\sigma}

\def\mmm{\mathcal{V}}
\def\mm2{\mathcal{W}}
\def\embed{\Psi}
\def\d{{\rm d}}
\def\const{\rm const.}

\newcommand{\bm}[1]{\mbox{\boldmath $#1$}}
\def\RR{\mathrm{I\hspace{-2.8pt}R}}

\def\FL{\mathrm{\scriptscriptstyle RW}}
\def\sx{\mathrm{\scriptscriptstyle SX}}
\def\stc{\mathrm{\scriptscriptstyle ST}}

\def\spfl{(\mmm^{\FL}, g^{\FL})}
\def\spsx{(\mmm^{\sx}, g^{\sx})}
\def\spst{(\mmm^{\stc},g^{\stc})}

\def\flr{(\mm2^{\FL}, g^{\FL})}
\def\sxr{(\mm2^{\sx}, g^{\sx})}
\def\str{(\mm2^{\stc}, g^{\stc})}

\def\slicefl{\mathcal{M}}

\def\tphi{\Phi}

\def\ax{\eta}
\def\axfl{\eta_{\FL}}
\def\st{\zeta}
\def\stk{\xi}

\def\no{\hat n}

\def\lie{\mathcal{L}}

\def\Journal#1#2#3#4#5#6{(#5) ``#6'' {#1} {\bf #2} #3#4}

\def\CQG{\em Class. Quantum Grav.}

\def\PRD{\em Phys. Rev. D }

 \def\IJT{\em Int. J. Theor. Phys.}
\def\PR{\em Phys. Rev.}
\def\RMP{\em Rev. Mod. Phys.}
 \def\MNRAS{\em Mon. Not. Roy. Astr. Soc.}
\def\JMP{\em J. Math. Phys.}
 \def\CMP{\em Commun. Math. Phys.}
\def\PRL{\em Phys. Rev. Lett.}
\def\AP{\em Ap. J.}

\newtheorem{theorem}{Theorem}
\newtheorem{proposition}{Proposition}
\newtheorem{lemma}{Lemma}
\newtheorem{coroprop}{Corollary}[proposition]

\newtheorem{definition}{Definition}
\newtheorem{remarkdef}{Remark}[definition]

\newtheorem{remarkpro}{Remark}[proposition]

\def\fin{\hfill \rule{2.5mm}{2.5mm}\\ \vspace{0mm}}
\def\finn{\hfill \rule{2.5mm}{2.5mm}}

\begin{document}
\title{Axially symmetric equilibrium regions of
Friedmann-Lema\^{\i}tre-Robertson-Walker universes.}
\author{Brien C. Nolan and Ra\"ul Vera}
\address{School of Mathematical Sciences, Dublin City
University, Glasnevin, Dublin 9, Ireland.}
\eads{\mailto{brien.nolan@dcu.ie, raul.vera@dcu.ie}}
\begin{abstract}
The study of the matching of stationary and axisymmetric
spacetimes with Friedmann-Lema\^{\i}tre-Robertson-Walker
spacetimes preserving the axial symmetry is presented. We show, in
particular, that any orthogonally transitive stationary and
axisymmetric region in FLRW must be static, irrespective of the
matter content. Therefore, previous results on static regions in
FLRW cosmologies apply. As a result, the only stationary and
axisymmetric vacuum region that can be matched to a (non-static)
FLRW spacetime is a spherically symmetric region of Schwarzschild.
This constitutes another uniqueness result for the Einstein-Straus
model (as well as its Oppenheimer-Snyder counterpart), and hence
another indication of its unsuitability as an answer to the
influence of the cosmic expansion on local physics.
\end{abstract}
\submitto{\CQG} \pacs{04.20.Jb, 04.20.Cv, 98.80.-k, 04.40.Nr}
\maketitle

\section{Introduction and summary}
In dealing with isolated systems in general relativity, one
usually assumes that the system is resident in an asymptotically
flat space-time. This is an idealisation, as the universe at large
is clearly not asymptotically flat. Recognising this fact, one is
then confronted with the issue of the impact of cosmology on local
physics. Ellis has written widely on this topic; see for example
the recent review \cite{ellis-nar}. Here and elsewhere he makes
the point that the perspective that cosmology can influence local
physics inverts the usual logical sequence: cosmology to a great
degree entails the generalisation of local physical laws to the
scale of the entire universe. Consider this example from black
hole physics. It has long been recognised \cite{penrose+simpson}
that a perturbation of compact support - created perhaps by the
collapsing object that forms the black hole - impinging upon the
inner (Cauchy) horizon of a charged spherical black hole undergoes
infinite blue-shift and ultimately converts the Cauchy horizon to
a scalar curvature singularity (see \cite{brady} for a review).
However such a black hole would be irradiated by the cosmic
microwave background radiation - a perturbation of non-compact
support. It was only recently that Burko \cite{burko} checked that
this perturbation of cosmic origin has the same effect as a
perturbation of local origin: the creation of a scalar curvature
singularity at the Cauchy horizon.

Asymptotically flat systems in general relativity can be studied
via the introduction of idealised structures (surfaces)
representing space-like and null infinity. These surfaces are
infinitely far away from the local system (e.g.\ star, binary
system) that one is studying. A central idea of Ellis' programme
for the study of the influence of cosmology on local physics is
that these surfaces should be replaced by surfaces at a finite
distance from the local system, representing what he refers to as
`finite infinity', and beyond which a cosmological model provides
the appropriate description of space-time. This surface should
have certain characteristics. For example, the gravitational field
encountered must be sufficiently small in some quantifiable way
and there should be limits on the radiation and matter content of
the surface guaranteeing that the system interior to the surface
is indeed isolated to a sufficiently high degree of approximation.
The point of view of the present paper is that a natural candidate
for the construction of finite infinity is a matching hypersurface
conjoining portions of two different space-times, one (the
interior) corresponding to the local system and the other (the
exterior) corresponding to a cosmological model.

This is not a new idea, nor is the idea that the cosmological
background may influence local systems. For example, soon after it
was discovered that the universe is expanding, McVittie addressed
the question of whether or not this expansion would influence
planetary orbits \cite{mcv}. His approach to this problem involved
determining an exact solution of Einstein's field equations that
represents the Schwarzschild solution embedded in a
Friedmann-Lema\^{\i}tre-Robertson-Walker (FLRW)
background, and then studying the geodesic
equations of this space-time. However, it is not clear that the
interpretation of this solution as representing a point mass
embedded in a Robertson-Walker background is entirely accurate, as
the resulting space-time has troubling global pathologies
\cite{brienmcv}. Furthermore, McVittie's analysis relies on a
flawed interpretation of a non-invariantly defined coordinate
radius as the radius of the planetary orbit (see Section 3.3 of
\cite{kras}). The first self-consistent and formally correct study
of this problem was given by Einstein and Straus
\cite{einstein-straus}. Here, they showed that the Schwarzschild
solution can be matched across a co-moving time-like
boundary of a dust-filled FLRW universe. The resulting structure is
referred to as the Einstein-Straus (ES) vacuole. An arbitrary
number of such vacuoles can be seeded in the cosmological
background, giving rise to the so-called Swiss Cheese model. As
regards planetary orbits, the conclusion is that there is no
influence from the cosmic background, as the planets move in the
familiar spherical vacuum.

However, there are many drawbacks associated with this model.
Principal among these are the following. First, exact spherical
symmetry of the vacuole is required. Second, the mass parameter of
the Schwarzschild vacuum region is directly related to the radius
of the vacuole and to the density of the FLRW background. This means
that the model is highly inflexible. See \cite{bonnor}. Thirdly,
the model is unstable against radial perturbations \cite{kras}.
Therefore, it is highly desirable to have at hand generalisations
of the ES model. Regarding the mass relation, a direct generalisation
exists by using a (radiative) Vaidya interior \cite{FST}. Nevertheless,
spherical symmetry still plays a fundamental role.

Unfortunately, the ES model has shown itself to be remarkably
reluctant to admit non-spherical generalisations. Attempts so far have
emphasised the equilibrium nature of the interior region, while
moving away from the constraint of spherical symmetry. Thus
Senovilla and Vera considered the matching of a static
cylindrically symmetric region with a FLRW universe and found that
this matching is impossible \cite{JRcyl}. Similarly, Mars
considered first static axially symmetric configurations
\cite{MARCaxi} and then {\em all} static configurations
\cite{MarcES} for a possible interior. The results of these
studies play a central role in the present paper and so are quoted
in detail in Section 5. These results can be summarised as
follows: at each instant of cosmic time, the static region
has a spherically symmetric boundary.
However, the centre of these spheres moves, and their radius change,
and so the overall configuration is not spherically symmetric in
general. In fact, the whole configuration is axially symmetric.
Furthermore, imposing a structure on the matter
distribution of the interior region (including the cases of vacuum
and perfect fluid) implies that full spherical symmetry is obtained.
In particular, the only static vacuole that may be embedded in a
FLRW universe is the spherically symmetric ES vacuole.

Given that, for reasons outlined above, one would like to be able
to embed into a FLRW background the vacuum gravitational field of a
non-spherical isolated system, and in particular, a non-spherical
isolated system in equilibrium, one must look for a get-out clause
to release us from the no-go results quoted above. Our attempt to
do this involves studying {\em stationary} rather than static
configurations. Thus rotation is allowed, but no gravitational
radiation. We study the standard model of a rotating isolated
system in equilibrium, the class of stationary axially symmetric
gravitational fields. We can express our results briefly and
prosaically as follows: it doesn't work. More precisely, we find
that \textit{the stationary region must in fact be static}, and so Mars'
results apply. In particular we can conclude that
{\em the only
stationary axially symmetric vacuum region that can be matched
with a FLRW universe must be spherically symmetric and is therefore
an ES vacuole}. The restrictions and instabilities mentioned above
then also apply.

The structure of this paper is as follows. In the next section, we
review the theory of matching two space-times across a general
(i.e.\ possibly character-changing) hypersurface \cite{MASEhyper}.
We use the separation from the full set of matching conditions of
the constraint matching conditions introduced by Mars
\cite{MARCaxi}. In Section 3, we describe in detail the setting of
the problem and give the required mathematical definitions. In
Section 4 we study the constraint matching conditions and
reproduce a result found by Mars in the static case, that the
matching hypersurface at each instant of time is a metric
2-sphere. In Section 5 the remaining matching conditions are
imposed and we prove our main result, that the stationary Killing
vector field must in fact be a static Killing vector field. Thus
the present case being studied becomes a case covered by the
results of Mars in \cite{MarcES}, and so we quote in full the
principal results of this work. We use the notation and
conventions of \cite{sol}.

\section{Summary on junction of spacetimes}
The formalism for matching two $C^2$ spacetimes\footnote{A $C^m$
spacetime is a Hausdorff, connected $C^{m+1}$
manifold with a $C^m$ Lorentzian metric (convention
$\{-1,1,1,1\}$).} $(\mm2^\pm,g^\pm)$ with respective boundaries
$\sup^\pm$ of arbitrary, and even changing, causal character was
presented in \cite{MASEhyper}.
The starting point for a further development
of the matching conditions by subdividing them into constraint and evolution
equations by using a 2+1 decomposition was introduced in \cite{MARCaxi}
(see also \cite{MarcES}). For completeness, let us devote this section for
a summary of the formalisms, and we refer to
\cite{MASEhyper,MARCaxi,MarcES} for further details.

Gluing $(\mm2^+,g^+,\sup^+)$ to $(\mm2^-,g^-,\sup^-)$ across their boundaries
consists of constructing a manifold $\mmm=\mm2^+\cup\mm2^-$ and
identifying both the points and the tangent spaces of $\sup^+$ and
$\sup^-$. This is equivalent to introducing
 an abstract three-dimensional $C^3$ manifold
$\suup$ and two $C^3$ embeddings $\embed_\pm$
\[
\embed_\pm~~ :~~ \suup~ \longrightarrow ~ \mm2^\pm
\]
such that $\embed_\pm(\suup)=\sup^\pm$. The identification of
points on $\sup^+$ and $\sup^-$ is performed by the diffeomorphism
$\embed_-\circ\embed_+^{-1}$, and we denote by $\sup(\subset
\mmm)\equiv  \sup^+ = \sup^-$ the identified hypersurfaces, i.e.
the matching hypersurface in $\mmm$. The conditions that ensure
the existence of a continuous metric $g$ in $\mmm$, such that
$g=g^+$ in $\mmm\cap\mm2^+$ and $g=g^-$ in $\mmm\cap\mm2^-$ are
the so-called \textit{preliminary junction conditions} and require
first the equivalence of the induced metrics on $\sup^\pm$, i.e.
\begin{equation}
  \label{eq:hs}
  \embed^*_+(g^+)=\embed^*_-(g^-),
\end{equation}
where $\embed^*$ denotes the pull-back of $\embed$. Secondly, one
requires the existence of two $C^2$ vector fields $\vec\ell_\pm$
defined over $\sup^\pm$, transverse everywhere (i.e.\ nowhere
tangent) to $\sup^\pm$, with different relative orientation
(by convention,
$\vec\ell_+$ points $\mm2^+$ inwards whereas $\vec\ell_-$ points
$\mm2^-$ outwards) and satisfying
\begin{equation}
  \label{eq:riggings}
  \embed^*_+(\bm \ell_+)=\embed^*_-(\bm \ell_-),\hspace{1cm}
  \embed^*_+(\bm \ell_+(\vec\ell_+))=
  \embed^*_-(\bm \ell_-(\vec\ell_-)),
\end{equation}
where $\bm\ell_\pm=g^\pm(\vec\ell_\pm,\cdot)$.
The existence of these so-called rigging vector fields is not ensured
when the boundaries have null points in some situations \cite{sign}.

Now, the Riemann tensor in $(\mmm,g)$ can be
defined in a distributional form \cite{MASEhyper}. In order
to avoid singular terms in the Riemann tensor
on 
$\sup$, a second set of conditions must be imposed. This second set
demands the equality of the so-called generalized second fundamental forms
with respect of the rigging one-forms, and can be expressed as
\begin{equation}
  \label{eq:second}
  \embed^*_+(\nabla^+\bm\ell_+)=\embed^*_-(\nabla^-\bm\ell_-),
\end{equation}
where $\nabla^\pm$ stands for the Levi-Civita covariant derivative
in $(\mm2^\pm,g^\pm)$. If conditions (\ref{eq:second}) are
satisfied for one choice of pair of riggings, then they do not
depend on the choice of riggings \cite{MASEhyper}. It must also be
stressed that although the so-called generalised second
fundamental forms $\mathcal{H}^{\pm}_{ab}\equiv
\embed^*_\pm(\nabla^\pm\bm\ell_\pm)_{ab}$, where $a,b\ldots=1,2,3$,
are not symmetric in general, the equations (\ref{eq:second}),
also denoted sometimes as $\mathcal{H}^+_{ab}=\mathcal{H}^-_{ab}$,
are indeed symmetric \cite{MASEhyper}.

Once the whole set of matching conditions hold, the finite one-side limits of
the Riemann tensor of $(\mmm,g)$ on $\sup$,
and in any $C^1$ coordinate system covering $\sup$ (or part thereof),
satisfy the following relation
\begin{equation}
  \label{eq:riemanns}
  \left.R^+_{\alpha\beta\mu\nu}\right|_{\sup}=\left.
  R^-_{\alpha\beta\mu\nu}+n_\alpha n_\mu B_{\beta\nu}
  -n_\beta n_\mu B_{\alpha\nu}-n_\alpha n_\nu B_{\beta\mu}+
  n_\beta n_\nu B_{\alpha\mu}\right|_{\sup},
\end{equation}
where $R^\pm_{\alpha\beta\mu\nu}$ are the Riemann tensors of
$(\mm2^\pm,g^\pm)$, respectively,
$\bm n$ is the normal
one-form to $\sup$, and $B_{\alpha\beta}$ is a symmetric tensor
which is defined up to the transformation
\begin{eqnarray*}
  B_{\alpha\beta}~\rightarrow~ B_{\alpha\beta}+X_\alpha n_\beta+
  X_\beta n_\alpha,
\end{eqnarray*}
for arbitrary one-form $\bm X$.

Following \cite{MARCaxi,MarcES},
the 2+1 splitting of the matching conditions starts by foliating
$(\suup,\embed^*_-(g^-))$ with a set of spacelike $C^3$ two-surfaces
$\suup_\tau$
where $\tau\in \RR$.
Let $i_\tau: \suup_\tau\rightarrow \suup$ be the inclusion map of
$\suup_\tau$ into
$\suup$. The compositions $\embed_{\tau,\pm}\equiv \embed_\pm\circ i_\tau$
define embeddings of $\suup_\tau$ into $(\mm2^\pm,g^\pm)$,
and the images $S^\pm_\tau\equiv \embed_{\tau,\pm}(\suup_\tau)$ are spacelike
two-surfaces lying on $\sup^\pm$ by construction.
Clearly, the identification of $\sup^+$ with $\sup^-$
through $\embed_-\circ\embed_+^{-1}$ induces the
identification of $S_\tau^+$ with $S_\tau^-$ by
the diffeomorphism $\embed_{\tau,+}\circ ({\embed_{\tau,-}})^{-1}$.
The identified surfaces will be denoted by $S_\tau\equiv S_\tau^+=S_\tau^-$,
and thus $S_\tau\subset\sup$.
For any given point $p\in S_\tau$, let us
denote by $N_p S^\pm_\tau$ the two-dimensional Lorentzian vector
space, subset of the cotangent space $T^*_p\mm2^\pm$,
spanned by the normal one-forms of $S^\pm_\tau$ at $p$.
The (normal) bundle with fibers $N_pS^\pm_\tau$ with base
$S^\pm_\tau$ will be denoted by $NS^\pm_\tau$.

The matching conditions impose restrictions on $S_\tau$ for each value
of $\tau$. These are called the
\textit{constraint matching conditions} and consist of two parts.
First, the restriction of the preliminary junction conditions on $S_\tau$
imposes the isometry of $S^+_\tau$ and $S^-_\tau$, i.e.
\begin{equation}
  \label{eq:preconstraint}
  \embed^*_{\tau,+}(g^+)=\embed^*_{\tau,-}(g^-).
\end{equation}
Secondly, and in order to ensure the identification of the tangent
spaces in $\sup^\pm$, for every $p\in S_\tau$ there must exist
a linear and isometric map
\begin{equation}
  \label{eq:isomap}
  f^p_\tau~:~ N_p S^+_\tau~\longrightarrow~ N_p S^-_\tau,
\end{equation}
with the following property, inherited from (\ref{eq:second}): the
second fundamental form of $S^+_\tau$ with respect to any section
$\bm m: S_\tau\to NS^+_\tau$, denoted by $\bm K^+_{S_\tau}(\bm m)\equiv
\embed^*_{\tau,+}(\nabla^+\bm m)$, and the corresponding image
through $f_\tau$, i.e. the normal one-form field $f_\tau(\bm m)$ to
$S^-_\tau$, will have to coincide, i.e.
\begin{equation}
  \label{eq:secconstraint}
  \bm K^+_{S_\tau}(\bm m)=\bm K^-_{S_\tau}(f_\tau(\bm m)),\hspace{1cm}
  \forall  \mbox{ sections } \bm m :S^+_\tau \to NS^+_\tau.
\end{equation}
For further details and more explicit form of the above expressions
we refer to \cite{MARCaxi}.

\section{Definitions and setting of the problem}
Regarding the FLRW spacetime, and since we will follow the procedures
used in \cite{MarcES}, let us review first some notation and conventions.
For completeness, we also use this section to review the definitions
and some assumptions involved in stationary and axisymmetric spacetimes.
\begin{definition}
Let $(\slicefl,g_\slicefl)$ be a complete, simply connected,
three-dimensional Riemannian manifold of constant curvature and let
$I\subset \RR$ be an open interval.
A FLRW spacetime $\spfl$ is the manifold $\mmm^{\FL}=I\times \slicefl$
endowed with the metric $g^{\FL}=-\d t\otimes \d t+a^2(t)g_\slicefl$, where
the so-called scale factor $a(t)$ is a positive $C^3$ function on $I$,
and such that
\begin{enumerate}
\item The energy density $\rho$ and the pressure of the cosmological
flow $p$ satisfy $\rho\geq 0$, $\rho+p\neq0$,
\item the expansion $3(\dot a /a) $ vanishes nowhere on $I$ (dot denotes
$d/dt$).
\end{enumerate}
\end{definition}
\begin{remarkdef}
{\em The energy ``condition'' $\rho\geq 0$ is automatically satisfied
whenever the constant curvature of $g_\slicefl$ is non-negative, and
it is used only in Proposition
\ref{res:first} with the effect of excluding some
spatially non-compact boundaries for the stationary and axisymmetric region.
Thus, this condition could be replaced by spatial compactness
of $\sup$. Requiring $\rho\geq 0$ is preferable as it holds
for most physically reasonable FLRW spacetimes.}
\end{remarkdef}
\begin{remarkdef}
{\em The condition $\dot a\neq 0$ is made for simplicity and
is not a fundamental restriction. The results presented here
apply to the expanding or contracting
disjoint regions of any non-static FLRW spacetime.}
\end{remarkdef}

We will denote by $\pi$ the canonical projection from $\mmm^{\FL}$
into $\slicefl$ and by $\slicefl_t$ the hypersurfaces
$\{t=\const\}$ in $\mmm^{\FL}$.

To start with, no specific matter content in the stationary and
axisymmetric region will be assumed, although the corresponding
$G_2$ on $T_2$ (necessarily) Abelian group \cite{commu} (see below) of
isometries
will be assumed to act orthogonally transitively (OT).

\begin{definition}\label{def:stax}
The OT stationary and axisymmetric
spacetime $\spsx$ is characterised by the existence
of a coordinate system $\{T,\tphi,x^M\}$
($M,N,...=2,3$) in which the line-element for the
metric $g^{\sx}$ outside the axis
reads \cite{sol}
\begin{equation}
  \label{eq:ds2sx}
  ds^2_{\sx}=-e^{2U}\left( d T+Ad\tphi\right)^2 +
e^{-2U}W^2 d \tphi^2 + g_{MN}d x^M d x^N,
\end{equation}
where $U$, $A$, $W$ and $g_{MN}$ are functions of $x^M$,
the axial Killing vector field is given by $\vec\ax=\p_{\tphi}$,
and a timelike (future-pointing) Killing vector field is given by $\vec\stk=\p_{T}$.
\end{definition}
\begin{remarkdef}
{\em The intrinsic definition of stationary and axisymmetric spacetime
consists of demanding
(i) that the spacetime admits a two-dimensional group of isometries
$G_2$ acting simply-transitively on timelike surfaces $T_2$
and containing a (spacelike) cyclic subgroup, so that $G_2=\RR\times S_1$,
and (ii) that the set of fixed points of the cyclic group is not empty.
Consequences of the definition are that
$G_2$ group has to be Abelian
\cite{commu},
and that the set of fixed points
must form a timelike two-surface \cite{maseaxconf}, this is the axis.
The axial Killing $\vec\ax$ is then intrinsically defined by
normalising it demanding
$\partial_\alpha\vec\ax^{\,2}\partial^\alpha\vec\ax^{\,2}/4\vec\ax^{\,2}\to 1$
at the axis.
See also \cite{sol} and \cite{jaumeax}.
}
\end{remarkdef}
\begin{remarkdef}
{\em The assumption of orthogonal transitivity on the group of
isometries is also known as the {\em circularity condition}, and
it turns out to be not an assumption, but a consequence of the
Einstein equations in most of the cases we will be interested in
eventually. Indeed, the $G_2$ on $T_2$ group must act orthogonally
transitively in a region that intersects the axis of symmetry
whenever the Ricci tensor has an invariant 2-plane spanned by the
tangents to the orbits of the $G_2$ on $T_2$ group
\cite{carter69}. By the Einstein equations, this includes
$\Lambda$-term type matter (cosmological constant), in particular
{\em vacuum}, perfect fluids without convective motions, and also
stationary and axisymmetric electrovacuum \cite{sol}.}
\end{remarkdef}
Allowed coordinate changes $\{x^\alpha\}\rightarrow \{\tilde x^\alpha\}$
keeping the form (\ref{eq:ds2sx})
with the axial Killing vector field reading $\vec\ax=\p_{\tilde\tphi}$
are given by
\[
  \tilde T = \alpha_0 T + \alpha_1, ~~
  \tilde \tphi = \tphi + \alpha_2 T +\alpha_3,~~~
  \tilde x^M = \tilde x^M(x^N),
\]
where the $\alpha$'s are constants.

Special attention is given to the one-form
\[\bm \st\equiv - D^2 \d T,\]
where we have defined
\begin{equation}
\label{eq:D}
D\equiv
\left[
\frac{g^{\sx}(\vec\ax,\vec\stk)^2}
     {g^{\sx}(\vec\ax,\vec\ax)}    -g^{\sx}(\vec\stk,\vec\stk)\right]^{1/2},
\end{equation}
and its corresponding vector field $\vec\st$,
\begin{equation}
\label{eq:defst}
\vec\st = \vec \stk -
\frac{g^{\sx}(\vec\ax,\vec\stk)}{g^{\sx}(\vec\ax,\vec\ax)} \vec\ax,
\end{equation}
which is orthogonal to the hypersurfaces of constant
$T$
as well as orthogonal to $\vec\ax$ by construction.
In
fact, $\vec\st$ is intrinsically defined as the future-pointing
timelike vector field tangent to the orbits of the $G_2$ group,
orthogonal to the axial Killing vector field \cite{bardeen1970}
(see \cite{sol}), and whose modulus is given by
$g^{\sx}(\vec\st,\vec\st)=-D^2$,
with the above intrinsically defined function $D$.
Note that $\vec\st$ is hypersurface
orthogonal, although it is not a Killing vector field in general.
The explicit expression of the norm of $\vec\ax$ in terms of
metric functions is $g^{\sx}(\vec\ax,\vec\ax)=W^2 D^{-2}$, where
$D=\left(e^{-2U}-A^2 W^{-2}e^{2U} \right)^{-1/2}$.
It must be stressed here that, first,
$D$ is indeed a real function, since the orbits of the $G_2$ group
are timelike. Secondly,
the ratio $g^{\sx}(\vec\ax,\vec\stk)
/g^{\sx}(\vec\ax,\vec\ax)$ must be finite on $\mmm^{\sx}$
(see \cite{jaumeax}).
And thirdly, that $D$, being then
well defined all over $\mmm^{\sx}$
(its value at the axis being
$[-g^{\sx}(\vec\xi,\vec\xi)]^{1/2}$), does not vanish anywhere.

We will denote by $\{\vec E_M\}$ any two linearly independent
vector fields spanning the surfaces orthogonal to the orbits of
the $G_2$ group, so that the set $\{\vec\st,\vec\ax,\vec E_M\}$
constitute a basis of the tangent spaces at every point in
$\spsx$ outside the axis. In the coordinate system used in
(\ref{eq:ds2sx}) one can simply take the choice $\vec
E_M=\p_{x^M}$.

The aim is to study the matching of a FLRW region to a (OT)
stationary and axisymmetric region across a common boundary. In
practice, the procedure consists of studying $C^3$ embeddings
$\embed^{\sx}:\suup \to \mmm^{\sx}$ and $\embed^{\FL}:\suup \to
\mmm^{\FL}$ such that $\sup^{\sx}\equiv \embed^{\sx}(\suup)$ and
$\sup^{\FL}\equiv \embed^{\FL}(\suup)$ are boundaries of
submanifolds-with-boundary $\mm2^{\sx}\subset \mmm^{\sx}$ and
$\mm2^{\FL}\subset \mmm^{\FL}$ respectively. As an aside, this
implies that $\sup^{\FL}$ and $\sup^{\sx}$ are embedded $C^3$
hypersurfaces without boundary.

Note that $\sup^{\sx}$ and $\sup^{\FL}$ split locally $\mmm^{\sx}$
and $\mmm^{\FL}$ into two complementary regions with boundary,
respectively, say $\mm2^{\sx}_1$ and $\mm2^{\sx}_2$, and
$\mm2^{\FL}_1$ and $\mm2^{\FL}_2$. Therefore, a priori, four
different possible scenarios arise from the combinations of
matchings of the $\mm2$ halves, although only two of them are
actually inequivalent, this is, if one matching is possible, then
its equivalent ``dual'' is also possible and both are solved
at the same time \cite{FST}. The choice between the two halves
that are going to be matched, simply denoted by $\mm2^{\sx}$ and
$\mm2^{\FL}$, is determined by the choice of the rigging vectors,
which will point outwards from $\mm2^{\sx}$ and inwards to
$\mm2^{\FL}$ by convention. The two spacetimes we want to match
are then $\sxr$ and $\flr$, to form a matched spacetime
$(\mmm,g)$, where $\mmm=\mm2^{\sx}\cup \mm2^{\FL}$, the matching
hypersurface is $\sup(\subset \mmm)\equiv \sup^{\sx}=\sup^{\FL}$
and the metric $g$ is $g^{\sx}$ in $\mm2^{\sx}$ and  $g^{\FL}$ in
$\mm2^{\FL}$.

Regarding the matching hypersurface $\sup$, the only assumptions
that will be made are, on the one hand, that it preserves the
axial symmetry \cite{mps}, present in $\sxr$ and, obviously, also
in $\flr$. This in short implies that $\sup$ is arranged so that
it is tangent to the trajectories defined by the axial Killing
vector field $\vec \ax$ on the $\sup^{\sx}$ side, and to any axial
Killing vector in FLRW, say $\vec\axfl$, on the $\sup^{\FL}$ side.

On the other hand, and following \cite{MarcES}, we will restrict
ourselves to the case in which $\sup$ is {\em generic}, this is,
such that the cosmological time on $\sup^{\FL}$ has no local
maximum or minimum. Non-generic hypersurfaces could be included in
the results, but at a high notational cost. For completeness, let
us include here the definitions involved as were presented in
\cite{MarcES}.

$\flr$ admits a privileged, future-pointing, timelike unit vector
field $\vec u=\p_t$ describing the velocity of the cosmological
flow. It is intrinsically defined as the normalised timelike
eigenvector of the Ricci tensor (pointing to the future) whenever
$\rho+p\neq 0$.
One defines now a map $\chi:\sup^{\FL}\to \RR$ which assigns to each
$x\in \sup^{\FL}$ the value of the cosmic time $t$ at $x$.
Since $\sup^{\FL}$ and $\mm2^{\FL}$ are $C^3$, it follows that
$\chi$ is a $C^3$ map. The Morse-Sard theorem implies then that the set
of critical values of $\chi$ has measure zero on $\RR$.
If the range of $\chi$ is itself of measure zero, and because
$\sup^{\FL}$ has no boundary, then $\sup^{\FL}$ is a collection
of hypersurfaces of constant time.
That case is of not interest for us, because
it does not describe a region inside (or outside) a FLRW spacetime.
As mentioned in \cite{MarcES}, this case is not difficult to study,
but to include it in the following treatment would make the
notation cumbersome. To avoid this situation and similar cases,
we will use the following definition \cite{MarcES}:

\begin{definition}
\label{def:generic}
A hypersurface $\sup^{\FL}$ in a FLRW spacetime is called {\em generic}
if and only if the function $\chi:\sup^{\FL}\to \RR$ defined above
has no local maximum or minimum on $\sup^{\FL}$.
\end{definition}
\begin{remarkdef}
{\em Geometrically, this means that for any point $p$
of a generic hypersurface and for any neighborhood $U$ of that point,
a portion of $U$ is in the future of $p$ and another portion
is in the past of $p$. Let us note that, as in \cite{MarcES},
we restrict ourselves to generic hypersurfaces for the sake of
simplicity in the notation, and that the general case
can be covered with little more effort from the results
found for generic hypersurfaces.
It must be stressed that a timelike
or null hypersurface is automatically generic.}
\end{remarkdef}
\begin{remarkdef}
{\em In many interesting cases, the matching conditions imply
that $\sup$ must necessarily be generic. In particular,
if the region matching FLRW is assumed to be {\em vacuum}, it is
well known that
$\sup$ must be tangent to $\vec u$, and therefore timelike everywhere.}
\end{remarkdef}

For a generic hypersurface $\chi(\sup^{\FL})$ is open in $I$. Let
us denote by $J$ the set of regular (i.e.\ non-critical) values of
$\chi$. By the implicit function theorem, the pre-image
\[
S^{\FL}_\tau\equiv \chi^{-1}(\tau)=\slicefl_\tau\cap \sup^{\FL}
\]
for any $\tau\in J$ is a two-dimensional $C^3$ embedded submanifold
of $\sup^{\FL}$. These surfaces foliate an open set
$\sup_o^{\FL}\subset \sup^{\FL}$. If $\sup^{\FL}$ is generic
then $\sup_o^{\FL}$ is dense in $\sup^{\FL}$.

\section{The constraint matching conditions}
We start by imposing the constraint matching conditions on the
foliation $\{S^{\FL}_\tau\}$ of $\sup_o^{\FL}$.
Let us
denote by $\sup^{\sx}_o$
the points in $\sup^{\sx}$ that correspond to $\sup^{\FL}_o$
under
the identification $\sup\equiv \sup^{\sx}=\sup^{\FL}$,
and simply by $\sup_o$ such identification,
i.e. $\sup_o\equiv \sup^{\sx}_o=\sup^{\FL}_o$.
For all $\tau\in J$ we define $\vec s$ as the unit normal
vector to $ S^{\FL}_\tau$
which is tangent to $\slicefl_\tau$ and points inwards in $\mm2^{\FL}$.
By construction, $\vec s$ is nowhere tangent to $\sup_o^{\FL}$.

The first step to solving the matching problem is given by the following
Proposition regarding the matching from the FLRW side,
the final result of which is analogous to that found in \cite{MarcES}
in the study of static regions inside FLRW.
For completeness, we also include the remarks made
by Mars in \cite{MarcES}.

\begin{proposition}
\label{res:first} Let $(\mmm,g)$ be the matching spacetime between
a FLRW region $\flr$ and an OT stationary and axisymmetric region
$\sxr$ across a connected, generic matching hypersurface $\sup$
preserving the axial symmetry. Let $S^{\FL}_\tau$ be the foliation
of $\sup_o^{\FL}$ as defined above.
Then, the following geometrical properties hold:
\begin{enumerate}
\item[a.]
The restriction of $\vec\st$ to $\sup_o$ is orthogonal to each surface
$S^{\FL}_\tau$ at any point $p\in S^{\FL}_\tau$. This implies that
there exists a function $T_o:J\subset\RR\to \RR$ such that the
surfaces $S^{\sx}_{\tau}\equiv \sup^{\sx}_o\cap\{T=T_o(\tau)\}$
must be identified with $S^{\FL}_\tau$ on $\sup_o$, i.e.
$S_\tau\equiv S^{\FL}_\tau=S^{\sx}_{\tau}$.
\item[b.] The hyperbolic angle between $\vec u$ and $\vec \st$ is a non-zero
constant on each connected component of the surface $S_\tau$.
\item[c.] Each connected component of $S^{\FL}_\tau$, and hence
$S_\tau$ and $S^{\sx}_{\tau}$, is a two-sphere
with the standard metric and it is an umbilical submanifold
in $(\mmm,g)$. Furthermore, there exists a spherically
symmetric coordinate system $\{t,r,\theta,\phi\}$ in
$\flr$ such that this surface corresponds to
$r=\const$ and $t=\const$
\end{enumerate}
\end{proposition}
\begin{remarkpro}
{\em Connectedness of $\sup$ is assumed for convenience. The local
nature of the matching conditions infers that the assumption
implies no loss of generality.
For an arbitrary $\sup$, Proposition \ref{res:first} holds
for any of its connected components.}
\end{remarkpro}
\begin{remarkpro}
{\em Conclusion (a) means, in other words, 
that the foliation by cosmic time and the foliation by the intrinsically
defined time $T$ must agree on the matching hypersurface.}
\end{remarkpro}
\begin{remarkpro}
{\em No topological assumptions are made on $\sup$ except for connectedness.
It is remarkable that spatial compactness follows from the matching
conditions.}
\end{remarkpro}
\begin{remarkpro}
{\em Conclusion (c) states that each surface $S_\tau$ is a
coordinate two-sphere. However, these spheres need not be
concentric with each other, that is, the center of each
$S^{\FL}_\tau$ in FLRW is still allowed to move arbitrarily with
cosmic time.}
\end{remarkpro}
\textit{Proof:} We start by fixing a regular value $\tau_0$ of $\chi$
and the
corresponding surface $S^{\FL}_{\tau_0}$.
For any point $p\in S^{\FL}_{\tau_0}$ consider an open neighbourhood
$U\subset \sup^{\FL}_o$ of $p$.
Let us denote by $\vec e_A$ ($A,B,C=1,2$) a pair of vector fields
on $U$ (restricting the size of $U$ if necessary)
which are linearly independent at every point and
tangent to the foliation $\{S^{\FL}_\tau\}$,
and define $h_{AB}=g(\vec e_A,\vec e_B)|_U$,
where $g(\cdot,\cdot)$ represents the scalar product in the matched
spacetime $(\mmm,g)$.
Note that $h|_{S^{\FL}_\tau}$ is the induced metric
on $S^{\FL}_\tau$.
To complete the basis of $T_q \mmm$ for every $q\in U$ we take the
restriction of the
fluid velocity vector on $U$, $\vec u|_U$, and the vector field $\vec s$,
defined on $U$ as mentioned above.
By construction, see above, $\vec u|_U$\footnote{Not to overwhelm the notation,
vectors $\vec v$  (and functions) defined only on $U$ will appear
as either $\vec v$ or the redundant expression $\vec v|_U$ in the
following expressions.} and $\vec s$ are mutually orthogonal,
and orthogonal to $\vec e_A$.
The set of vectors $\{\vec u,\vec s,
\vec e_{A}\}$ constitute then a basis of $T\mmm$ on $U$.
Since $\vec s$ is transverse to
$\sup^{\FL}_o$, we have $\bm n(\vec s)\neq 0$, where $\bm{n}$ is
normal to $\sup^{\FL}$.
By the 
identification of $\sup^{\sx}$ and
$\sup^{\FL}$ in $\sup\subset \mmm$, the vector field $\vec \st$
at any $q\in U$ can be expressed in the basis
$\{\vec u,\vec s, \vec e_{A}\}$ as
\begin{equation}
  \label{eq:stU}
  \left.\vec\st\right|_{U}=
  \left. D\cosh\beta\, \vec u -D\sinh\beta\cos\alpha \vec s
  +c^A \vec e_A\right|_{U},
\end{equation}
where 
$\alpha$, $\beta$, $c^A$ are
$C^2$ scalar functions on $U$ and $c^A$ satisfy
\begin{equation}
\label{eq:cs}
  c^A c^B h_{AB}=D^2\sinh^2\beta\sin^2\alpha.
\end{equation}

Because of the preservation of the axial symmetry
\cite{mps},
the restriction of $\vec\ax$ to
$\sup$ is tangent to $\sup$,
as well as the restriction of an axial
Killing vector field in $(\mm2^{\FL},g^{\FL})$, say $\vec\axfl$.
Furthermore, by their intrinsic (global) characterisations,
$\vec\ax$ and $\vec\axfl$ must be identified on $\sup$ so that
$\mmm$ admits a (continuous) axial symmetry.
Since $\vec\axfl|_{\sup}$ 
is necessarily also tangent to
the foliation $\{S^{\FL}_\tau\}$, all this means
\begin{equation}
  \label{eq:axU}
  \left.\vec\ax\right|_U=\left.\vec\axfl\right|_U=
  \left. \ax^A \vec e_A\right., 
\end{equation}
for some functions $\ax^A$ defined on $U$.
Mutual orthogonality of $\vec\st$ and $\vec\ax$ demands that
\begin{equation}
  \label{eq:orth1}
  c^A \ax^B h_{AB}=0.
\end{equation}
It can be easily checked that the following two vector
fields defined on $U$,
\begin{equation}
  \label{eq:vs}
  \left.\vec v_A \right. 
  \equiv \left.\left[\bm n
    \left(\vec s-\tanh\beta\,\cos\alpha\,\vec u\right)\right]\vec e_A+
  \frac{h_{AB}c^B}{D\cosh\beta}
  \left[\bm n(\vec s)\vec u -\bm n(\vec u) \vec s\right]\right|_{U},
\end{equation}
are tangent to $\sup$ and orthogonal to $\vec\st$.
Computing
\[
g(\vec v_A,\vec\ax)|_{U}=\left.\left[\bm n
    \left(\vec s-\tanh\beta\cos\alpha\vec u\right)\right]
  h_{AB} \eta^B\right|_{U}
\]
shows that the vector
\begin{equation}
  \label{eq:defw}
  \vec v= c^A \vec v_A
\end{equation}
on $U$, apart from being tangent
to $\sup$ and orthogonal to $\vec\st$ by construction, is also orthogonal
to $\vec\ax$ by virtue of (\ref{eq:orth1}).
Therefore, there exist two functions $v^M$ on $U$
such that $\vec v=v^M\vec E_M|_U$, (see (\ref{eq:ds2sx})).

The Riemann tensor in FLRW reads
\begin{eqnarray*}
R^{\FL}_{\alpha\beta\mu\nu}&=& \frac{\varrho+p}{2}\left[
 u_\alpha u_\mu g^{\FL}_{\beta\nu}
-u_\alpha u_\nu g^{\FL}_{\beta\mu} +u_\beta u_\nu
g^{\FL}_{\alpha\mu} -u_\beta u_\mu g^{\FL}_{\alpha\nu} \right]\\
&&+\frac{\varrho}{3} \left(
 g^{\FL}_{\alpha\mu}g^{\FL}_{\beta\nu}
-g^{\FL}_{\alpha\nu}g^{\FL}_{\beta\mu}
\right).
\end{eqnarray*}
Due to the orthogonal transitivity
in the stationary and axisymmetric region, it is straightforward
to show that the following identities hold
\begin{equation}
  \label{eq:guais}
  R^{\sx}_{\alpha\beta\mu\nu} \stk^\alpha \ax^\beta \ax^\mu E^\nu_M=0,~~~
  R^{\sx}_{\alpha\beta\mu\nu} \st^\alpha E^\beta_M E^\mu_N E^\nu_O=0.
\end{equation}
As a result, the contraction of (\ref{eq:riemanns})
with $\st^\alpha$, $\ax^\beta$, $\ax^\mu$ and $v^\nu$ on $U$
reads
$
  0=-(1/2)(\varrho+p)~g(\vec u,\vec \st)~ g(\vec u,\vec v_A)c^A
  g(\vec\ax,\vec\ax) 
  |_{U},
$
which by virtue of (\ref{eq:vs}) and (\ref{eq:stU}) can be expressed as
\begin{equation}
  \label{eq:main1}
  0=\left.\frac{\varrho+p}{2}~\bm n(\vec s)g(\vec\ax,\vec\ax)\,
    h_{AB} c^A c^B ~
  \right|_{U}.
\end{equation}
Using $\varrho+p\neq 0$, the fact that $h_{AB}$ is positive definite
and that $\vec\ax$
only vanishes at points in the axis,
one has that $c^A$ must vanish on a dense subset of $U$,
and thus $c^A=0$ 
on $U$ by continuity.
Therefore, by making use of (\ref{eq:cs}), expression (\ref{eq:stU}) becomes
\begin{equation}
  \label{eq:stUf}
  \left.\vec\st\right|_{U}=\left.
    D\cosh\beta\, \vec u - D\sinh\beta\, \vec s \right|_{U}.
\end{equation}
As a result, the slicing on $U$ defined by $S^{\sx}_{\tau}\equiv
\sup^{\sx}_o\cap\{T=T_o(\tau)\}$ for some function
$T_o:J\to \RR$ must be identified with
$S^{\FL}_\tau{}$ via the diffeomorphism induced by the matching
procedure, this is $(S_{\tau}\equiv) S^{\sx}_\tau=S^{\FL}_{\tau}$,
and conclusion (a) follows.

Expression (\ref{eq:stUf}) tells us that $\bm \st|_q\in
N_qS^{\sx}_\tau$ for every $q\in U$, and it can be reexpressed in
the language of (\ref{eq:isomap}), using $g^{\FL}$ to lower the
indices of $\vec u$ and $\vec s$,
as
\begin{equation}
  \label{eq:fst}
  f^q_\tau(\bm \st|_q)=\left. D\cosh\beta~ \bm u - D\sinh\beta~ \bm s\right|_q
\end{equation}
for every $q\in U$.
It is also convenient to introduce the vector
$\vec\lambda$, defined as being tangent to the foliation $\{S_\tau\}$,
orthogonal to $\vec\ax|_{\sup}$ and with its same modulus,
and oriented so that on $U$ has the form
\[
\vec\lambda=\epsilon^{AB}h_{BC}\ax^C \vec e_A\equiv \lambda^A\vec e_A,
\]
where $\epsilon^{AB}=-\epsilon^{BA}$, $\epsilon^{12}=1$.
Since it is also orthogonal to $\vec\stk$ by construction,
then it will have the form $\vec\lambda=\lambda^M \vec E_M|_{\sup}$
as seen from $\sup^{\sx}$.
The contractions
of the second fundamental form of $S^{\sx}_{\tau_0}$ with
respect to $\bm \st$, 
i.e. $\bm K^{\sx}_{S_{\tau_0}}\left(\bm
\st|_{S^{\sx}_{\tau_0}}\right)_{AB}$,
which is symmetric,
with the vectors
$\{\vec\lambda,\vec \ax\}$
read
\begin{eqnarray}
  \label{eq:kst}
  &&
  \bm K^{\sx}_{S_{\tau_0}}\left(\bm \st|_{p}\right)_{AB}
  \lambda^A \lambda^B|_p=
  \bm K^{\sx}_{S_{\tau_0}}\left(\bm \st|_{p}\right)_{AB}
  \ax^A \ax^B|_p=0,\nonumber\\
  &&\bm K^{\sx}_{S_{\tau_0}}\left(\bm \st|_{p}
  \right)_{AB}\ax^A \lambda^B|_p=
  -\left.\frac{1}{2}\,g(\vec\eta,\vec\eta)\, \vec\lambda
  \left(\frac{g(\vec\ax,\vec\stk)}{g(\vec\ax,\vec\ax)}\right)
\right|_{p},
\end{eqnarray}
for every $p\in S^{\sx}_{\tau_0}$.
On the other hand, on the FLRW side,  one has
\begin{equation}
  \label{eq:covu}
  \nabla^{\FL}_\alpha u_\beta=\frac{\dot a}{a}
  \left(g^{\FL}_{\alpha\beta}+ u_\alpha u_\beta\right),
\end{equation}
and thence the second fundamental form
of $S^{\FL}_{\tau_0}$ with respect to $\bm u$ at any
$p\in S^{\FL}_{\tau_0}$ reads
\begin{equation}
\label{eq:Kforu}
\bm K^{\FL}_{S_{\tau_0}}\left(\bm u|_{p}\right)_{AB}=
\left.\frac{\dot a}{a} h_{AB}\right|_p.
\end{equation}
Regarding $\bm s$, the crucial point here is that, because
of the preservation of one isometry across $\sup$,
which is an axial isometry in FLRW,
the second fundamental
form $\bm K^{\FL}_{S_{\tau_0}}\left(\bm s|_{S^{\FL}_{\tau_0}}\right)$
is diagonal in the basis
$\{\vec\lambda,\vec\ax\}$.
Indeed, since $\{\vec \lambda,\vec \ax\}$ span the surfaces
$S_\tau$ (except at the axis, where they vanish),
which are orthogonal to $\vec s$, we have
$g^{\FL}(\vec s,[\vec\lambda,\vec\ax_{\FL}])|_{S^{\FL}_{\tau}}=0$.
On the other hand, since $\vec\axfl$ is an axial
Killing vector field of FLRW, it is hypersurface orthogonal,
and thus $\bm \axfl\wedge \d \bm\axfl=0$,
which means $\d \bm\axfl=\bm \mu\wedge\bm\axfl$
for some one-form $\bm\mu$.
As a result, the following chain of equalities hold:
\begin{eqnarray}
\bm K^{\FL}_{S_{\tau_0}}\left(\bm s|_{p}\right)_{AB}
\lambda^A \ax^B|_p&=&
\left.\lambda^\alpha \axfl^\beta\nabla^{\FL}_\alpha s_\beta
\right|_{p}
=-\left.\lambda^\alpha s^\beta\nabla^{\FL}_\alpha {\axfl}_\beta
\right|_{p}\nonumber\\
&=&-\left.\lambda^\alpha s^\beta\nabla^{\FL}_{[\alpha}
{\axfl}_{\beta]} \right|_{p}\nonumber\\&=& \left.\lambda^\alpha
s^\beta\left(
\mu_\beta{\axfl}_\alpha-\mu_\alpha{\axfl}_\beta\right)\right|_{p}=0,
\label{eq:offdiag}
\end{eqnarray}
for every $p\in S^{\FL}_{\tau_0}$.

We are now ready to apply the constraint matching equations
(\ref{eq:secconstraint}) to $\bm \st|_{S^{\sx}_{\tau_0}}$
using (\ref{eq:fst}).
In other words, we are going to equate the second fundamental
vector form of $S_\tau$ with respect to $\bm\st$
as computed from the $S^{\sx}_\tau$ side
to that computed from the $S^{\FL}_\tau$ side:
\begin{equation}
  \label{eq:constraintexpl}
  \bm K^{\sx}_{S_\tau}\left(\bm \st|_{S^{\sx}_\tau}\right)_{AB}=
  \bm K^{\FL}_{S_\tau}
  \left(\left.D \cosh\beta~ \bm u - D\sinh\beta~ \bm s\right|_{S^{\FL}_\tau}
    \right)_{AB}.
\end{equation}
Due to (\ref{eq:offdiag}),
the non-diagonal part of (\ref{eq:constraintexpl}),
i.e. the contraction of (\ref{eq:constraintexpl}) with $\ax^A,\lambda^B$,
leads to the vanishing
of (\ref{eq:kst}) for every $p\in S^{sx}_{\tau_0}$,
which implies
\begin{equation}
  \label{eq:nodiag}
\left.
  g(\ax,\ax)\,
  \vec\lambda
  \left(\frac{g(\vec\ax,\vec\stk)}{g(\vec\ax,\vec\ax)}\right)
\right|_{S^{\sx}_{\tau_0}}=0.
\end{equation}
Summing up, so far the constraint matching conditions have lead us to
(see (\ref{eq:kst}))
\begin{equation}
  \label{eq:constraintmc}
  \bm K^{\sx}_{S_{\tau_0}}\left(\bm \st|_{S^{\sx}_{\tau_0}}\right)=0.
\end{equation}
By virtue of (\ref{eq:constraintmc}) and (\ref{eq:fst}),
the constraint matching conditions (\ref{eq:constraintexpl})
imply now
$\bm K^{\FL}_{S_{\tau_0}}\left(\bm u|_{S^{\FL}_{\tau_0}}\right)-
\tanh\beta\, \bm K^{\FL}_{S_{\tau_0}}\left(\bm s|_{S^{\FL}_{\tau_0}}\right)=0$.
Since $\dot a$ is nowhere zero by assumption, this relation
together with (\ref{eq:Kforu}) imply that $\beta$ is nowhere
zero on $S^{\FL}_{\tau_0}$, and therefore
\begin{equation}
  \label{eq:ks}
  \bm K^{\FL}_{S_{\tau_0}}\left(\bm m|_{S^{\FL}_{\tau_0}}\right)=
  \left.\frac{\dot a}{a} ~
  \bm m\left(-\vec u + \coth \beta \vec s\right)~
  h\right|_{S^{\FL}_{\tau_0}}
\end{equation}
for every normal one-form $\bm m$ 
to $S^{\FL}_{\tau_0}$.
Equation (\ref{eq:ks}), in particular, tells us that $S^{\FL}_{\tau_0}$ is
umbilical in FLRW, and hence in the resulting matched spacetime
$(\mmm,g)$.

At this point, the rest of the proof follows strictly the proof of
Proposition 1 in \cite{MarcES}. Nevertheless, and for completeness
regarding point (2), let us include here the expression of the
Riemann tensor of the induced metric $h$ of $S^{\FL}_{\tau_0}$
(and thus of $S_{\tau_0}$), which is obtained from (\ref{eq:ks})
by using the Gauss' equation (see e.g. \cite{MASEhyper}):
\[
R^{(2)}_{ABCD}(\tau_0)=
\left.\left(\frac{\rho}{3}+\frac{\dot a^2}{a^2}\frac{1}{\sinh^2\beta}\right)
(h_{AC} h_{BD}- h_{AD} h_{BC})\right|_{S^{\FL}_{\tau_0}}.
\]
Using $\rho\geq 0$, it follows that $S^{\FL}_{\tau_0}$ has positive
constant curvature, and hence it is locally {\em isometric} to a two-sphere
(with the standard metric).
In particular, because $S^{\FL}_{\tau_0}$ is umbilical
in the maximally symmetric space $\mathcal{M}_{\tau_0}$,
by using the Codazzi equation it is straightforward
to show that $\beta$ is constant along each connected component
of $S^{\FL}_{\tau_0}$. This proves conclusion (b).
As mentioned, the reader is referred to \cite{MarcES}
for the proof of conclusion (c).\fin

From Proposition \ref{res:first}, so far
we have that 
the matching hypersurface $\sup_o$ is formed by a collection of
two-spheres moving on space with arbitrarily changing radius. The
assumption that $\sup^{\FL}$, and hence $\sup$, be generic,
together with the differentiability assumed ($C^3$) also prevents
these spheres from merging. 
Indeed, if two
spheres merge by touching each other in one point first, then
$\sup$ would cease to be $C^3$. Alternatively, if two spheres
merge at all points simultaneously, then $\chi$ should present a
maximum (or a minimum) there, in contradiction with having a
generic embedded $\sup$.

Let us consider now
any connected component of $\sup\setminus \sup_o$,
and denote it by $S_{c}$.
Following the same argument as above, since $\sup$ is $C^3$ and
$\sup_o$ is foliated by two-spheres, the only possibility left
for any $S_{c}$ is to be a 
two-sphere itself.
This is, the level surfaces corresponding to the critical values of $\chi$,
i.e. the critical levels $S_c$,
are also two-spheres.
One can therefore extend the definition of the vector field $\vec s$
to all $\sup^{\FL}$ at the critical level surfaces $S_{c}$
by continuity. Then, clearly, equations
(\ref{eq:stUf}) and (\ref{eq:ks}) are valid for all $\tau \in I$.

This is summarised as follows:

\begin{coroprop}
Let $\sup^{\FL}$ be the hypersurface in Proposition \ref{res:first}.
Then
$\{S^{\FL}_\tau\}$, where now
$S^{\FL}_\tau\equiv \slicefl_\tau\cap \sup^{\FL}$
for all $\tau\in I$, defines a foliation on the whole of $\sup^{\FL}$,
and the three points of Proposition \ref{res:first} hold
for all $S_\tau$.
\end{coroprop}
In particular, when $\sup$ is connected, then each element of the foliation
$\{S_\tau\}$ is connected as well.


It must be stressed that the analogous
Corollary in \cite{MarcES} presents the erroneous further statement
that no critical points of $\chi$ can exist. Nevertheless,
that has no consequences for the final results, as the features
we are interested in on
$\sup_o$ extend to $\sup$ by continuity. Also, although the final
equations describing the matching hypersurface in FLRW and the
form of the static metric as presented in \cite{MarcES}
are valid only on $\sup_o$ and its neighbourhoods, they can be
found to apply as well on $S_{c}$ by taking a limit.
As with the study of non-generic hypersurfaces, the critical
points of $\chi$ in the generic hypersurface $\sup$ could be incorporated
to the results, but at the expense of introducing some more notation.
Since our aim is to find qualitative features for the
stationary region, and these are found to be necessary
around a dense subset of $\sup$, they will have to hold
all over $\sup$ and its neigbourhood and thus the more explicit
behaviour of $\spsx$ at the critical values is irrelevant.

\section{The rest of the matching conditions}
So far we have only used the constraint matching conditions, a
subset of the matching conditions, and we have found Proposition
\ref{res:first} as a result.
We are now ready to apply the full set of matching conditions.
As mentioned, for simplicity in the presentation,
we will not consider the critical points of $\sup$,
and thus we will analyse the matching conditions on $\sup_o$.

In order to analyse the remaining
conditions we need first
to construct 
a basis for $T\sup$ and then an explicit expression for a rigging
vector to be used in the equations (\ref{eq:second}).
We start by taking two vectors $\{\vec e_A\}$
spanning $S_\tau$ locally, as in the proof of Proposition
\ref{res:first}.
To complete the basis for $T\sup$ we choose now a
third vector 
tangent to $\sup$ and orthogonal to
$\{\vec e_A\}$. 
At points on $\sup_o$, we will denote that vector as
$\vec m$, and since $g(\vec u,\vec m)\neq 0,$
we choose it such that $g(\vec m,\vec u)|_{\sup_o}=-1$
for convenience.
By definition, $\vec m$ 
can be explicitly decomposed
(as seen from $\sup^{\FL}$) as
\begin{equation}
\label{eq:mwithmu}
\vec m|_{\sup_o}= \vec u + \mu \vec s|_{\sup_o},
\end{equation}
where $\mu$ is a $C^2$ function defined on $\sup_o$.
Since we have $\bm n(\vec s)\neq 0$ in $\sup_o$, we can choose
a normal one-form $\bm \no$ defined on $\sup_o$
so that $\bm \no(\vec s)=1$ there,
and thus
it decomposes as
\begin{equation}
\label{eq:normal}
\bm \no\equiv \mu\bm u + \bm s|_{\sup_o}
\end{equation}
in the $\{\vec u,\vec s, \vec e_{A}\}$ basis.
Note that $\bm \no$ is not unit, and that we will keep denoting
by $\bm n$ any normal to $\sup$. In order to find
the expression for $\vec m$ in the $\sup^{\sx}$ side we still need
to define a basis of $T\mmm$ on $U$ in relation to the
$\sup^{\sx}$ side, to compare with the orthogonal basis $\{\vec
u,\vec s, \vec e_{A}\}$ previously constructed in relation to
the $\sup^{\FL}$ side. For that we simply need a vector $\vec w$
orthogonal to $\vec \lambda$, $\vec\ax$ and $\vec\st$,
and with norm $D^2$,
to complete an orthogonal set $\{\vec\st,\vec\eta,\vec\lambda,\vec w\}$
at points on $\sup$,
constituting a basis of $T\mmm$ outside the axis.
By construction
and by virtue of (\ref{eq:stUf}),
we have that $\vec w$ (up to a sign) decomposes as
\[
\vec w=w^M \vec E_M|_{\sup}=
-D\sinh\beta\, \vec u+ D\cosh\beta\,\vec s|_{\sup},
\]
for some two $C^2$ functions defined on $\sup$, $w^M$,
which have to satisfy $\lambda^N w^M g^{\sx}(\vec E_M,\vec E_N)=0$
and $w^N w^M g^{\sx}(\vec E_M,\vec E_N)=D^2$.
In this basis, 
$\vec m$ can be expressed as
\begin{equation}
\label{eq:m}
  \vec m=D^{-2}\left[\bm \no(\vec w) \vec\st-\bm \no(\vec \st) \vec w\right],
\end{equation}
for
\begin{equation}
\label{eq:nwnx}
\bm \no(\vec w)=D(\mu\sinh\beta+\cosh\beta)|_{\sup_o},~~~
\bm \no(\vec \st)=-D(\mu\cosh\beta+\sinh\beta)|_{\sup_o}.
\end{equation}

In considering a rigging vector for the
computation of the remaining matching conditions,
the following fundamental result will simplify things.
In the following, we indicate with a prime
the sets to which points of the axis have been removed,
this is $A'\equiv \{x\in A;\vec\ax(x)\neq 0\}$.
\begin{lemma}
\label{res:riggings} The vector $\vec \st$ is nowhere tangent to
$\sup'$ (and neither is $\vec w$). Consequently, the
stationary Killing vector field $\vec\xi$ is nowhere tangent to
$\sup'$. 
\end{lemma}
\textit{Proof:}
Computing the contraction of the identity (\ref{eq:riemanns})
with $\st^\alpha,\ax^\beta,\ax^\mu,w^\nu$, recalling
that $\vec w=w^M \vec E_M|_{\sup}$, and using
the geometrical identities (\ref{eq:guais}) that hold on $\mm2^{\sx}$,
we obtain
\begin{equation}
  \label{eq:ids_saaw}
  \bm n(\vec\st)\, \bm n(\vec w) B_{\alpha\beta}\,
  \ax^\alpha \ax^\beta|_{\sup}=
  \left.\frac{\rho+p}{2}\,
    D^2\cosh\beta\sinh\beta\, g(\vec\ax,\vec\ax)\right|_{\sup}.
\end{equation}
Note that this equation is valid all over $\sup$.
The tensor $B_{\alpha\beta}$ is at least continuous on $\sup$,
$\beta$ is nowhere zero on $\sup$ (from Proposition \ref{res:first},
and its corollary).
We must thus have $\bm n(\vec\st)\neq 0$ (and  $\bm
n(\vec w)\neq 0$) everywhere in $\sup$ except, possibly, where $\vec\ax=0$.
The first statements of
the Lemma follow. The last statement trivially follows by
construction; (\ref{eq:defst}) $\Rightarrow\bm n(\vec\st)=\bm
n(\vec\stk)$.\fin
\\
This Lemma states, in other words,
that the matching hypersurface must be locally non-stationary
nearly everywhere.
This fact will be fundamental
in order to determine the geometry of the stationary and axisymmetric
region
by extending the constraints implied by the matching conditions on $\sup'$
into $\mm2'^{\sx}$, and thereby arrive at the final result.
In fact, although Lemma \ref{res:riggings}
leaves out of consideration the points at the axis,
the final result will eventually lead to the fact
that $\vec\st$ must be indeed transverse to the whole of $\sup$.

Let us study then the equations (\ref{eq:second})
by taking $\bm \ell=\bm \st|_{\sup}$.
The fact that $\vec\st$  {\em may} fail to be transverse to $\sup$
at points on the axis will not be of relevance.
Taking into account
(\ref{eq:stU}), conditions (\ref{eq:second}) read now
$
\embed^*_{\sx}(\nabla^{\sx}\bm\st|_{\sup})=
\embed^*_{\FL}(\nabla^{\FL}
(D\cosh\beta\, \bm u - D\sinh\beta\, \bm s|_{\sup})).
$
Denoting by $\{\vec e_a\}$
for $a,b,\ldots = 1,2,3$ a basis of $T\sup$ on $\sup$,
these equations read 
\begin{equation}
\label{eq:secondab}
\left.e_a{}^\alpha e_b{}^\beta\nabla^{\sx}_\alpha\st_\beta\right|_{\sup}=
\left.e_a{}^\alpha e_b{}^\beta\left[
 \nabla^{\FL}_\alpha (D\cosh\beta\, u_\beta) -
 \nabla^{\FL}_\alpha (D\sinh\beta\, s_\beta)\right]\right|_{\sup}.
\end{equation}
The antisymmetric part
of these equations must be satisfied once the preliminary junction conditions
(\ref{eq:hs})-(\ref{eq:riggings}) hold \cite{MASEhyper},
although depending on the procedure followed in the matching,
these might still give us information about implicit consequences
of (\ref{eq:hs})-(\ref{eq:riggings}).
The second set of matching conditions consist then
on the symmetric part in $(a,b)$ of (\ref{eq:secondab}),
and thereby provide six, in priciple independent, conditions.
Following the above construction, we can simply take
$\{\vec e_a\}=\{\vec m,\vec e_A\}$
on a neigbourhood $V\subset\sup$ (small enough to ensure that $\{\vec e_A\}$
are linearly independent there).
Notice that the $(A,B)$ components of (\ref{eq:secondab})
correspond to the constraint matching conditions (\ref{eq:constraintexpl}),
which have been already used.
The remaining matching conditions in (\ref{eq:second})
correspond then to the $(\vec m,\vec e_A)$ and
$(\vec m,\vec m)$ components of (\ref{eq:secondab}).

To compute those components of (\ref{eq:secondab}) we need the
expressions of the covariant derivatives of $\bm u$ and $\bm s$ on
$\sup$ in the basis $\{\vec u,\vec s,\vec e_A\}$. The first
has been already given in (\ref{eq:covu}),
and the second was given
in Lemma 2 in \cite{MarcES}. For completeness, we include it here
with a little more detail.
But first, let us note that in order to have defined a covariant
derivative of $\bm s$, as it is, one has to define an extension
of $\bm s$ off $\sup$, and that $\nabla_\alpha s_\beta$ depends
on that extension, in principle.
Despite that, we will
be eventually only interested on
$e_a{}^\alpha e_b{}^\beta \nabla^{\FL}_\alpha s_\beta $
(for (\ref{eq:secondab})),
which is independent of the extension. In fact,
no extension of $\bm s$ is needed for its calculation,
but for the sake of simplicity,
we prefer to present the covariant derivate 
for a particular extension.
Without loss of generality we can assume then,
as far as the following
lemma is concerned, that $\bm s$ (and thus also $\bm n$)
is extended off $\sup$ being unit and orthogonal to $\bm u$ and
a family of spheres concentric to $S^{\FL}_\tau$.
\begin{lemma}
The covariant derivative of $\bm s$ at points on $U\subset\sup_o$ reads
\[
\left.\nabla^{\FL}_\alpha s_\beta\right|_{U}
=
\left.\frac{\dot a}{a} s_\alpha u_\beta
+\frac{\dot a}{a}\coth\beta\,h^{AB} e_{A\alpha} e_{B\beta}
+u_\alpha h^{AB} \vec e_A(\mu) e_{B\beta}\right|_{U}.
\]
\end{lemma}
{\em Proof:}
Take $p\in \sup_o$ and a neighbourhood $U\subset V\cap\sup_o$
(this is $U$ as defined in Proposition \ref{res:first}).
Since $\vec s$ is unit, orthogonal to $\vec u$ and (\ref{eq:ks})
holds, then the covariant derivative of $\bm s$ decomposes as
\begin{equation}
\label{eq:covs}
\left.\nabla^{\FL}_\alpha s_\beta\right|_{\sup}=
\left.\frac{\dot a}{a} s_\alpha u_\beta
+\frac{\dot a}{a}\coth\beta\, \mathcal{X}_{\alpha\beta}
+u_\alpha v_\beta+ s_\alpha b_\beta
\right|_{\sup},
\end{equation}
where $\mathcal{X}_{\alpha\beta}=g_{\alpha\beta}+u_\alpha u_\beta-
s_\alpha s_\beta$ is the projector orthogonal to $\{\vec u,\vec s\}$,
and the vectors $\vec v$ and $\vec b$ are tangent to $\{S^{\FL}_{\tau}\}$.
Given any Killing vector field $\vec \kappa$
we have $s^\alpha \kappa^\beta\nabla^{\FL}_\alpha s_\beta=
\vec s\,\left(g^{\FL}(\vec\kappa,\vec s)\right)$.
Therefore,
for any Killing vector field $\vec\kappa$ orhogonal to $\vec s$,
we have $s^\alpha\kappa^\beta \nabla^{\FL}_{\alpha} s^{ }_{\beta}=0$,
which in view of (\ref{eq:covs}),
translates onto
$g^{\FL}(\vec\kappa,\vec b)=0$.
Since this has to hold
for any Killing vector $\vec \kappa$ generator of the $SO(3)$ isometry group
which $S^{\FL}_\tau$ and a family of concentric spheres
indeed admit, we thus have $\vec b=0$.
On $V$ we can clearly write $\mathcal{X}_{\alpha\beta}=
h^{AB} e_{A\alpha} e_{B\beta}$.
To determine $\vec v$ we use the fact that the normal one-form
$\bm n$ (in fact, its extension) is integrable,
so that, in particular at points on $\sup_o$, $\bm \no\wedge \d \bm \no=0$.
Using (\ref{eq:normal}), (\ref{eq:covs}) and $\d \bm u=0$
this equation becomes $\bm s \wedge \bm u \wedge (\d \mu-\bm v)=0$
at points in $U$.
This implies $\d \mu=\bm v +r_1 \bm u + r_2 \bm s$ for some
scalar functions $r_1$ and $r_2$,
and hence $v_A\equiv \bm v (\vec e_A)=\vec e_A(\mu)$.
Recalling that $\vec v$ is tangent to $S^{\FL}_\tau$, $\bm v$ decomposes
as $v_\alpha=v_A h^{AB} e_{B \alpha}$ and hence
$v_\alpha=h^{AB}\vec e_A(\mu) e_{B\alpha}$.\finn
\\

We are ready to compute
the $(\vec m,\vec e_A)$ and $(\vec m,\vec m)$
components in (\ref{eq:secondab}). Nevertheless, and
and as we have been doing previously,
we replace the basis $\{\vec e_A\}$ of $TS_\tau$ 
for the orthogonal pair $\{\vec\lambda,\vec\ax\}$.
The symmetrised contractions of (\ref{eq:secondab})
using $(\vec m,\vec\lambda)$, $(\vec m,\vec\ax)$ and
$(\vec m,\vec m)$ can be computed now on $\sup_o$ using (\ref{eq:mwithmu})
to obtain
\begin{eqnarray}
   \left.\bm \no(\vec w)\vec\lambda(D)\right|_{\sup_o}
   &=&
   D^2\sinh\beta\, \vec \lambda(\mu)|_{\sup_o}, \label{eq:ml}\\
   -g(\vec\ax,\vec\ax)
   \left.\vec m\left(\frac{g(\vec\ax,\vec\stk)}{g(\vec\ax,\vec\ax)}\right)
   \right|_{\sup_o}
   &=&
   D\sinh\beta\, \vec \ax(\mu)|_{\sup_o},\label{eq:me}\\
   0
   &=&
   \left.\bm \no(\vec w)\,\vec m(D)+
   D\,\bm \no(\vec\st)\left(\mu\frac{\dot a}{a}-\vec m(\beta)\right)\right|_{\sup_o},
   \label{eq:mm}
\end{eqnarray}
where we have used the fact that for scalar functions $f$
invariant under $\{\vec\stk,\vec\ax\}$, we have $\vec\st(f)=0$,
and thus $\bm \no(\vec\st)\vec w(f)= -D^{2}\vec m(f)$ by
(\ref{eq:m}).

At this point
we recall the fact that $\sup$, and hence
$\sup^{\FL}$, preserves the axial
symmetry, which ensures that
\begin{equation}
  \label{eq:mpsmu}
  \vec\ax(\mu)|_{\sup_o}=0.
\end{equation}
This can be shown to follow from the invariance of $g(\vec m,\vec
m)|_{\sup_o}=-1+\mu^2$. However for the sake of clarity and to
keep the analysis self-contained,
we present a proof 
using the previous construction. For $\vec\ax$ being a
Killing vector orthogonal to $\vec m$, we must have $m^\alpha
\ax^\beta\nabla^{\FL}_\alpha m_\beta|_{\sup}=0$. On the other hand,
using (\ref{eq:mwithmu}) 
and the above expressions of
the covariant derivatives for $\bm u$ and $\bm s$, we have
$m^\alpha \ax^\beta\nabla^{\FL}_\alpha
m_\beta|_{\sup_o}=-\mu\vec\ax(\mu)|_{\sup_o}$.
The combination of both equations
leads to (\ref{eq:mpsmu}).\footnote{
In fact, (\ref{eq:mpsmu}) is equivalent to the antisymmetrised
part of the $(\vec m,\vec\eta)$ component of (\ref{eq:secondab}).
Furthermore, it can be also checked that the antisymmetrised
part of the $(\vec m,\vec\lambda)$ component of (\ref{eq:secondab})
coincides with (\ref{eq:ml}), and thus (\ref{eq:secondab})
is exhausted with (\ref{eq:ml})-(\ref{eq:mpsmu}).}

The importance of (\ref{eq:mpsmu}) is that equations
(\ref{eq:ml})-(\ref{eq:mm}) eventually decouple to equations for
the FLRW side, namely (\ref{eq:ml}) and (\ref{eq:mm}) (with
(\ref{eq:mpsmu})), and one equation for the stationary
axisymmetric side, (\ref{eq:me}). Let us concentrate first on the
FLRW side.

Defining now on $\sup_o$
the function 
$\Delta\equiv \bm \no(\vec w)|_{\sup_o}=
D^{-1}(\mu\sinh\beta+\cosh\beta)|_{\sup_o}$,
equation
(\ref{eq:ml}) is equivalent to $\vec\lambda(\Delta)|_{\sup_o}=0$.
At the same time,
and since $\vec\ax(D)|_{\sup}=0$ and $\vec\ax(\beta)|_{\sup}=0$
($\beta$ in constant on every $S_\tau$ from Proposition \ref{res:first}
and its corollary),
together with (\ref{eq:mpsmu}), we have $\vec\ax(\Delta)|_{\sup_o}=0$.
On the other hand,
since $\bm n(\vec w)\neq 0$ everywhere in
$S'_\tau$ (from Lemma \ref{res:riggings}),
$\Delta$ cannot vanish anywhere on $\sup_o$..
(This, in turn, implies that
$\bm n(\vec w)$ cannot vanish anywhere in $\sup_o$ either.)

Furthermore, rewriting
equation (\ref{eq:mm}) for the function $Z\equiv \mu/a$ and using
the fact that $\vec s(a)=0$, so that
$\vec m(a)|_{\sup_o}=\dot a|_{\sup_o}$, we
arrive at the same result as Lemma 2 in \cite{MarcES}, which is
the final key result for the determination of the matching
hypersurface on the FLRW side:
\begin{lemma}
Given the conditions obtained in Proposition \ref{res:first}
for $\sup^{\FL}_o$,
the matching conditions (\ref{eq:second}),
as far as the FLRW side is concerned,
are satisfied on $\sup^{\FL}_o$ if and only if
$D|_{\sup_o}=\Delta^{-1}(\mu\sinh\beta+\cosh\beta)|_{\sup_o}$,
where $\Delta$ is a non-zero contant on each two-sphere $S^{\FL}_{\tau}$,
and the following partial differential equation on $\sup_o$
holds
\begin{equation}
  \label{eq:eqforZ}
  \vec m(Z)=Z^2 \dot a \coth\beta+
  Z\left[\frac{\vec m(\Delta)}{\Delta}-2\coth\beta\, \vec m(\beta)\right]
  +\frac{\vec m(\Delta)}{a\Delta}\coth\beta-\frac{2}{a}\vec m(\beta),
\end{equation}
where $Z\equiv \mu/a$.\fin
\end{lemma}
This Lemma is used in \cite{MarcES} to completely
and explicitly compute
the form of $\sup^{\FL}_o$ in some explicit coordinates,
which will be presented below.

Up to this point, we have shown that the embedding of OT
stationary and axisymmetric regions in FLRW spacetimes and the
embedding of static regions in FLRW spacetimes are equivalent as
far as the FLRW side is concerned. We obtain a stronger result
when analysing the implications on the $\mm2^{\sx}$ side. In short,
the stationary and axisymmetric region will necessarily be static,
and thus this case will come under the scope of the results found
in \cite{MarcES}.

\section{The stationary and axisymmetric region must be static}
We concentrate now on the consequences of the matching conditions
on the OT stationary and axisymmetric side.
The only conditions from (\ref{eq:second}) which we have not
analysed yet are those involving quantities on the $\mm2^{\sx}$
side. These are (\ref{eq:nodiag}), which is part of the constraint
matching conditions and holds everywhere on $\sup_o$ outside the
axis, and (\ref{eq:me}) taking (\ref{eq:mpsmu}) into account. In
short, and extending by continuity at critical points of $\chi$, we have
\[
g(\vec\ax,\vec\ax)\,\vec\lambda
\left.\left(\frac{g(\vec\ax,\vec\stk)}{g(\vec\ax,\vec\ax)}\right)
\right|_{\sup}=0,
~~~
g(\vec\ax,\vec\ax)\, \vec m
\left.\left(\frac{g(\vec\ax,\vec\stk)}{g(\vec\ax,\vec\ax)}\right)
\right|_{\sup}=0,
\]
which, together with the fact that $\vec\ax$ is a Killing vector,
imply that
$\Omega\equiv g^{\sx}(\vec\ax,\vec\stk)/g^{\sx}(\vec\ax,\vec\ax)$
is constant on $\sup'^{\sx}$,
and therefore everywhere on $\sup^{\sx}$ by continuity.
This result allows us first to extend Lemma \ref{res:riggings}
to all $\sup$.
\begin{proposition}
\label{res:riggings2} The vector $\vec \st$ is nowhere tangent to
$\sup$. Consequently, the
stationary Killing vector field $\vec\xi$ is nowhere tangent to
$\sup$. 
\end{proposition}
\textit{Proof:} Recalling that $\vec m(\Omega)|_{\sup}=0
\iff \vec w(\Omega)|_{\sup}=0$,
from the above discussion we have $\vec\lambda(\Omega)|_{\sup}=
\vec w(\Omega)|_{\sup}=0$. Furthermore, since $\vec\lambda$ is
tangent to $\sup$, then we also have
$\vec\lambda(\vec\lambda(\Omega))|_{\sup}=
\vec\lambda(\vec w(\Omega))|_{\sup}=0$.
Using these four equations on $\sup$,
a straightforward calculation shows that
$\st^\alpha \lambda^\beta \ax^\mu w^\nu
R^{\sx}_{\alpha\beta\mu\nu}|_{\sup}=0$
and 
$\st^\alpha \ax^\beta \lambda^\mu w^\nu
R^{\sx}_{\alpha\beta\mu\nu}|_{\sup}=0$.
Recalling that $\vec w=w^M \vec E_M|_{\sup}$
and $\vec \lambda=\lambda^M \vec E_M|_{\sup}$,
the geometrical identities in (\ref{eq:guais})
imply $\st^\alpha \lambda^\beta \lambda^\mu w^\nu
R^{\sx}_{\alpha\beta\mu\nu}|_{\sup}=0$ and
$\st^\alpha \ax^\beta \ax^\mu w^\nu
R^{\sx}_{\alpha\beta\mu\nu}|_{\sup}=0$.
Taking into account the decomposition of $\vec\lambda$ and
$\vec\ax$ in the basis $\{\vec e_A\}$ on $V\subset\sup$ defined around
any point $p\in\sup$,
the four previous equations read
$\st^\alpha {e_A}^\beta {e_B}^\mu w^\nu
R^{\sx}_{\alpha\beta\mu\nu}|_{q}=0$ at any $q\in V'$.
By continuity, that extends to any $q\in V$, and thus
\begin{equation}
  \label{eq:riestatic}
  \st^\alpha {e_A}^\beta {e_B}^\mu w^\nu R^{\sx}_{\alpha\beta\mu\nu}|_V=0.
\end{equation}
Computing now the contraction of the identity (\ref{eq:riemanns})
with $\st^\alpha,{e_A}^\beta,{e_B}^\mu,w^\nu$ on $V$,
and using (\ref{eq:riestatic}) we finally obtain
\[
  \bm n(\vec\st)\, \bm n(\vec w) B_{\alpha\beta}\,
  {e_A}^\alpha {e_B}^\beta|_{V}=
  \left.\frac{\rho+p}{2}\,
    D^2\cosh\beta\sinh\beta\, h_{AB}\right|_{V}.
\]
By the same argument as in Lemma \ref{res:riggings}
we must thus have $\bm n(\vec\st)|_{V}\neq 0$,
and hence $\bm n(\vec\stk)|_{V}\neq 0$.\fin

This means that the stationary Killing vector
$\vec\stk$ in $\spsx$ is transverse to
the hypersurface $\sup^{\sx}$ everywhere.
Using Lie-transport, this allows us to determine the
geometry of the $\spsx$ spacetime at least in a neighbourhood of
$\sup^{\sx}$. Although the main result, namely that the values of
$\Omega|_{\sup^{\sx}}$ extend on $\mmm^{\sx}$
is quite intuitive and immediate, for the sake of rigorousness and 
to present Mars' results,
let us introduce here all the machinery needed
for the extension and the notation used in \cite{MarcES}.

Let us define by $\Xi$ the space of the orbits of $\vec\stk$, this
is, the quotient space $\mmm^{\sx}/( \mbox{orbits of }\vec\stk)$,
so that $\mmm^{\sx}=I_1\times \Xi$ and $\vec\stk$ is tangent to
the $I_1\subset\RR $ factor. Basically, one can choose
$\Xi\equiv\{T=\const\}$. Define $\Pi:\mmm^{\sx}\to \Xi$ as the
canonical projection along the orbits of $\vec\stk$. Since
$\vec\stk$ is everywhere transverse to $\sup^{\sx}$, the
restriction of $\Pi$ on $\sup^{\sx}$ is a diffeomorphism between
$\sup^{\sx}$ and $\Pi(\sup^{\sx})\subset \Xi$. In particular, any
orthonormal basis at $p\in\sup^{\sx}$ can be uniquely extended to
an orthonormal tetrad on $\mathcal{U}\equiv
I_1\times\Pi(\sup^{\sx})$ by Lie transport along $\vec\stk$.
Following the same construction we have
used all throughout the paper, we take
$\{\vec e_A\}$ to be a pair of orthonormal
vector fields tangent to each $S^{\sx}_\tau$,
which are defined everywhere except for a pair of antipodal
points.
As usual, in order to cover those points we would need two patches,
but this is standard and we do not discuss it further.
The orthonormal tetrad for every point in $\sup^{\sx}$ can be now
completed by using $\vec e_0=D^{-1}\vec\st$, $\vec e_1=D^{-1}\vec
w$ and $\vec e_A$. Its dual frame will be denoted by $\{\bm
\theta^\alpha\}$. This tetrad can be extended to
$\mathcal{U}\subset\mmm^{\sx}$ by Lie transport along $\vec\stk$.

Having this in mind, we can finally prove the main result:
\begin{theorem}
\label{res:mainhere}
Assume that the OT stationary and axisymmetric spacetime $\spsx$
can be matched to a FLRW spacetime through a generic hypersurface
$\sup^{\sx}$ preserving the axial symmetry.
Then, the spacetime $\spsx$ must be, in fact, static
on a neighborhood $\mathcal{U}\subset\mmm^{\sx}$
of $\sup^{\sx}$. 
\end{theorem}
{\em Proof:} We have seen that the matching conditions
(\ref{eq:second}) imply that $\Omega\equiv
g^{\sx}(\vec\ax,\vec\stk)/g^{\sx}(\vec\ax,\vec\ax)$ is constant on the
hypersurface $\sup^{\sx}$ (i.e.\ $\Omega|_{\sup^{\sx}}=\const$),
to which $\stk$ is transverse, by Proposition
\ref{res:riggings2}. Let us consider the neighborhood $\mathcal{U}$
of $\sup^{\sx}$ previously defined. Since $\vec\stk$ is a Killing
vector we have $\lie_{\stk}\Omega=\vec \stk(\Omega)=0$; that is,
$\Omega$ is invariant under the action of $\vec\stk$ and thus is
constant along the trajectories of $\vec\stk$. Therefore
$\Omega|_{\mathcal{U}}=\Omega|_{\sup}=$ const.
As a result, the
vector $\vec\st=\vec\stk-\Omega\vec\ax$, (see (\ref{eq:defst})),
which is hypersurface orthogonal by construction, is also a
(timelike) Killing vector field on $\mathcal{U}$, that is, a
static Killing vector field.\fin

In a neighbourhood of the matching hypersurface,
the OT stationary axisymmetric spacetime $\spsx$
now becomes a static axisymmetric spacetime $\spst$ and the
results of Mars \cite{MarcES} apply, which we inculde
here for completeness. Without loss of
generality, the metric $g^{\stc}$ can be taken to be the metric
(\ref{eq:ds2sx}) with $A\equiv0$.

It must be stressed, though, that the results in \cite{MarcES}, as presented,
apply only on $\sup_o$, in principle,
since the existence of critical points of $\chi$
cannot be ruled out. Nevertheless, only an isolated critical levels
$S_c$ (if any) would remain to be treated.
In fact, the only point that would need a special treatment
corresponds to the equation (\ref{eq:eqforZ}),
and still, most of the results
can be obtained by considering the relations below at
a convenient limit $\mu\to\infty$.

The following result concerns the FLRW side, and determines
the most general form of $\sup^{\FL}_o$.
\begin{lemma}{\rm [Mars \cite{MarcES}]}
\label{res:supfl}
There exists a coordinate
system $\{t,r,\theta,\phi\}$ in the FLRW spacetime in which the hypersurface
$\sup_o$ is given by $r=r_0(t)$, where $r_0$ is $C^3$ and non-negative,
while the
line element for $g^{\FL}$ can be written as
\begin{eqnarray}
 ds^2 &=&-dt^2+a^2(t)\left[(dr+f(t)\cos\theta
dt)^2\right.\nonumber\\ &&
\left.+(\Sigma(r,\epsilon)d\theta-f(t)\Sigma_{,r}(r,\epsilon)\sin\theta
dt)^2+\Sigma^2(r,\epsilon)\sin^2\theta d\phi^2\right],
\label{rwlineel}\end{eqnarray} where
\[
\Sigma(r,\epsilon)=\left\{\begin{array}{ccc}
                             \sin r & {\rm{if}} & \epsilon=1, \\
                             r & {\rm{if}} & \epsilon=0, \\
                             \sinh r & {\rm{if}} & \epsilon=-1, \\
                           \end{array}
\right.
\]
and the function $f(t)$, together with
a further auxiliar function $\Delta(t)$,
satisfy the ordinary differential equations
\[ \dot{f}=X_1f,\qquad
f^2\left.\frac{\Sigma_{,r}}{\Sigma}\right|_{r=r_0(t)}+X_2=0,\]
where $X_1(t)$ and $X_2(t)$ are defined as
\begin{eqnarray*}
X_1&\equiv&\frac{2\epsilon_1\dot{a}\dot{r_0}}{\tanh\beta}+
\frac{\dot{\Delta}}{\Delta}-\frac{2\cosh\beta}{\sinh\beta}\dot{\beta},\\
X_2&\equiv&-\ddot{r_0}+\frac{\epsilon_1\dot{a}\dot{r_0}^2}{\tanh\beta}+
\frac{\dot{\Delta}}{\Delta}
\left(\dot{r_0}+\frac{\epsilon_1}{a\tanh\beta}\right)-
2\left(\frac{\epsilon_1}{a}+\frac{\dot{r_0}}{\tanh\beta}\right)\dot{\beta},
\end{eqnarray*}
for $\epsilon_1=\pm 1$ and $\beta(t)$ given
by
\[
\tanh\beta
=\epsilon_1
\left.
\frac{\Sigma\dot{a}}{\Sigma_{,r}}\right|_{r=r_0(t)}.
\]\finn
\end{lemma}
The hypersurface $\sup^{\FL}_o$ is represented then
by spheres whose centers (at $r=0$) move parallel to the axis
of symmetry defined by $\cos\theta=0$ (see \cite{MarcES}).
The curve $\sigma(t)$ describing that movement satisfies
$\dot\sigma(t)=f(t)$.

Regarding the geometry of the static region, the first
result concerns the Einstein tensor there, and the second,
the explicit geometry of the static region, which is the main
theorem in \cite{MarcES}.
\begin{proposition}{\rm [Mars \cite{MarcES}]}
\label{res:marc1} Consider the static space-time $\spst$
that can be matched to a FLRW spacetime across a generic boundary
$\Sigma$. Then there exists a neighbourhood $\mathcal{U}\subset\mm2^{\stc}$ of $\Sigma_o$ (as defined above) in the
matched spacetime on which the metric $g^{\stc}$ has the following
properties.
\begin{enumerate}
\item The energy-momentum tensor $T^{\stc}$ of $g^{\stc}$
decomposes as
\[
T^{\stc}=\rho^{\stc}\theta^0\otimes\theta^0+
p_r^{\stc}\theta^1\otimes\theta^1+
p_t^{\stc}(\theta^2\otimes\theta^2+\theta^3\otimes\theta^3),
\]
where the scalars
$\rho^{\stc},p_r^{\stc}$ and $p_t^{\stc}$ are given by
\begin{eqnarray*}
&&\rho^{\stc}=\frac{\rho\mu-p\tanh\beta}{\mu+\tanh\beta},\qquad
p_r^{\stc}=\frac{p-\rho\mu\tanh\beta}{1+\mu\tanh\beta},\\
&&p_t^{\stc}=\frac{3p-\rho}{6}+
\frac{\tanh\beta(2h^{AB}\vec{e}_A(\mu)\vec{e}_B(\mu)-
(\rho+p)(\mu^2-1)}{2(\mu+\tanh\beta)(1+\mu\tanh\beta)}\\
&&~~~~+\left[-\ddot{\beta}+\frac{\dot{\Delta}}{\Delta}
\left(\frac{\dot{a}}{a}\frac{1}{\tanh\beta}+\dot{\beta}\right)
-2\dot{\beta}\frac{\dot{a}}{a}\left(1+\frac{\mu}{\tanh\beta}\right)\right.\\
&&~~~~\left.+\mu\left(\frac{\ddot{a}}{a}+\mu\frac{\dot{a}^2}{a^2\tanh\beta}\right)
\right]\left(\cosh^2\beta(\mu+\tanh\beta)(1+\mu\tanh\beta)\right)^{-1},
\end{eqnarray*}
where \[\rho\equiv3\frac{\dot{a}^2+\epsilon}{a^2},\qquad
p\equiv-\frac{2a\ddot{a}+\dot{a}^2+\epsilon}{a^2}\] are the density and
pressure of the FLRW space-time.
\item The Petrov type of $g^{\stc}$ is $D$, the double principal
null directions are $\vec{e}_0\pm\vec{e}_1$ and the only
non-vanishing Weyl spin coefficient in the null basis canonically
associated with $\{\vec{e}_\alpha\}$ is
\[\Psi_2=\frac16\left(p-p_t^{\stc}-\frac{(\mu^2-1)(\rho+p)\tanh\beta}{(1+\mu\tanh\beta)(\mu+\tanh\beta)}\right).\]\finn
\end{enumerate}
\end{proposition}
\begin{theorem}{\rm [Mars \cite{MarcES}]}
\label{thm:marc2} Let $\spfl$ be a FLRW
spacetime and let $\spst$ be a static spacetime. Let
$\flr$ be an open submanifold of $\spfl$ with $C^3$
connected boundary $\sup^{\FL}$ and let $\str$ be an open
submanifold of $\spst$ with $C^3$ connected boundary $\sup^{\stc}$.
Assume that $\sup^{\FL}$ is generic
(according to Definition \ref{def:generic}).

Then a $C^0$ spacetime $(\mmm,g)$ can be constructed by gluing
$\mm2^{\FL}$ and $\mm2^{\stc}$ across their boundaries if and only if the
following conditions are satisfied.

\begin{enumerate}
\item
In the coordinate system of (\ref{rwlineel}), the boundary
$\sup^{\FL}$ of $\mm2^{\FL}$ is defined by the embedding
$\{t,\theta,\phi\}\to\{t,r=r_0(t),\theta,\phi\}$, where $r_0(t)$
is a $C^3$ non-negative function. The submanifold $\mm2^{\FL}$ is
defined by $\{\epsilon_1r\geq\epsilon_1r_0(t)\}$, where
$\epsilon_1=\pm1$.
\item
There exists a coordinate system $\{T,t,\theta,\phi\}$ in an
open neighbourhood $\mathcal{U}$ of $\sup^{\stc}$ in $\mm2^{\stc}$ such that
the static line element takes the form
\begin{eqnarray*}
ds^2&=&-\frac{(\cosh\beta+\mu\sinh\beta)^2}{\Delta^2}dT^2+(\mu\cosh\beta+\sinh\beta)^2dt^2\\
&&+a^2(t)\Sigma^2(r_0(t),\epsilon)\left[\left(d\theta-f(t)\left.\frac{\Sigma_{,r}}{\Sigma}\right|_{r=r_0(t)}\sin\theta
dt\right)^2+\sin^2\theta d\phi^2\right],\end{eqnarray*} where
$\Sigma(r,\epsilon), f(t), \beta(t)$ and $\Delta(t)$ are as defined
in Lemma \ref{res:supfl}
and $\mu\equiv\epsilon_1a(\dot{r_0}+f(t)\cos\theta)$.
\item
The boundary $\sup^{\stc}$ is defined in the $\{T,t,\theta,\phi\}$
coordinate system by the embedding\linebreak
$\{t,\theta,\phi\}\to\{T=T(t),t,\theta,\phi\}$ where $T(t)$
satisfies $\dot{T}(t)=\Delta(t)$.\finn
\end{enumerate}
\end{theorem}

Of course, imposing conditions on the matter content on the
(OT) stationary and axisymmetric (and hence static) region,
the use of Proposition \ref{res:marc1} restricts the possibilities
for the different functions that determine the matching hypersurface
and the static geometry. Indeed, let us consider a vacuum
stationary and axisymmetric region $\spsx$ matched to FLRW
preserving the axial symmetry.
As mentioned in the second remark in Definition \ref{def:stax},
orthogonal transitivity is not an assumption in this case.
Therefore Theorem \ref{res:mainhere} implies that the vacuum region
must be static. Then, Proposition \ref{res:marc1} applies, and, by
imposing $\rho^{\stc}=p_r^{\stc}=p_t^{\stc}=0$, a straightforward
calculation \cite{MARCaxi} gives $\mu=p=0$ and $f=0$, in particular.
The second remark in Definition \ref{def:generic} comes clear
in this case, since $\mu=0$ on $\sup_o$ and so it has to be
all over $\sup$ by continuity, and therefore $\sup=\sup_o$.
The vanishing of $f$
implies that the whole of the static region (not just its boundary)
has to be spherically symmetric,
and hence Schwarzschild,
by using Theorem \ref{thm:marc2}.
The Einstein-Strauss model is thus recovered
if the Schwarzschild region is inside FLRW and
the Oppenheimer-Snyder model \cite{OPSN} if it is outside.
This is summarised as follows:
\begin{theorem}
Let $(\mmm,g)$ be a spacetime resulting from matching a
FLRW region $\flr$ to a stationary and axisymmetric vacuum
region $(\mm2^v,g^v)$ across a $C^3$ connected matching hypersurface $\sup$
preserving the axial symmetry.
Then, the vacuum region $(\mm2^v,g^v)$ must be
a submanifold isometric to Schwarzschild, bounded by
a hypersurface formed by concentric spheres following
geodesics in FLRW.
\end{theorem}

In fact, for most reasonable energy-momentum tensors
in the stationary and axisymmetric region
(see second remark of Definition \ref{def:stax}), Theorem \ref{res:mainhere}
together with Proposition \ref{res:marc1}
will force the stationary and axisymmetric metric
and its boundary to be static and spherically symmetric.
The reader is referred to \cite{MARCaxi} for a detailed
discussion regarding the consequences of Proposition \ref{res:marc1}.

\ack We thank Marc Mars and Jos\'e Senovilla for discussions and
for suggesting numerous improvements to the manuscript.  We
acknowledge the algebraic manipulation computer CLASSI built on
SHEEP for some of the calculations. RV acknowledges the Irish
Research Council for Science, Engineering and Technology
postdoctoral fellowship PD/2002/108.


\section*{References}
\begin{thebibliography}{99}
\bibitem{ellis-nar}Ellis G F R \Journal{New Astron.
Rev.}{46}{645}{-657}{2002}{Cosmology and local physics}

\bibitem{penrose+simpson}Simpson M and Penrose R
\Journal{\IJT}{7}{183}{-197}{1973}{Internal instability in a
Reissner-Nordstrom black hole}

\bibitem{brady} Brady P R \Journal{Prog. Theor. Phys.
Suppl.}{136}{29}{-44}{1999}{The internal structure of black holes}

\bibitem{burko}Burko L \Journal{\PRD}{66}{024046}{}{2002}{Survival of the black hole's Cauchy horizon under non-compact
perturbations}

\bibitem{mcv}McVittie G C \Journal{\MNRAS}{93}{325}{-339}{1933}{The
mass-particle in an expanding universe}

\bibitem{brienmcv} Nolan B C \Journal{\PRD}{58}{064006}{}{1998}{A
point mass in an isotropic universe: existence, uniqueness and
basic properties}\newline Nolan B C
\Journal{\CQG}{16}{1227}{-1254}{1999}{A point mass in an isotropic
universe: II. Global properties}\newline Nolan B C
\Journal{\CQG}{16}{3183}{-3191}{1999}{A point mass in an isotropic
universe: III. The region $R\leq 2m$.}

\bibitem{kras}Krasi\'nski A (1997) {\em Inhomogeneous Cosmological
Models} (Cambridge: Cambridge University Press)

\bibitem{einstein-straus}Einstein A and Straus E G
\Journal{\RMP}{17}{120}{-124}{1945}{The influence of the expansion
of space on the gravitational fields surrounding the individual
stars}

\bibitem{bonnor}Bonnor W B
\Journal{\MNRAS}{282}{1467}{-1469}{1996}{The cosmic expansion and
local dynamics}

\bibitem{FST}Fayos F, Senovilla J M M and Torres R \Journal{\PRD}{54}
{4862}{-4872}{1996}
{General matching of two spherically symmetric spacetimes}

\bibitem{JRcyl} Senovilla J M M and Vera R
\Journal{\PRL}{78}{2284}{-2287}{1997}{Impossibility of the
cylindrically symmetric Einstein-Straus model}

\bibitem{MARCaxi}Mars M \Journal{\PRD}{57}
{3389}{-3400}{1998} {Axially symmetric Einstein-Straus models}

\bibitem{MarcES}Mars M \Journal{\CQG}{18}
{3645}{-3663}{2001} {On the uniqueness of the Einstein--Strauss
model}

\bibitem{MASEhyper}Mars M and Senovilla J M M \Journal{\CQG}{10}
{1865}{-1897}{1993}
{Geometry of general hypersurfaces in spacetime: junction conditions}

\bibitem{sol}Stephani H, Kramer D, MacCallum M A H,
Hoenselaers C and Herlt E (2003)
{\em Exact solutions of Einstein's field equations. Second Edition}, Cambridge
University Press, Cambridge

\bibitem{sign}Mars M, Senovilla J M M and Vera R \Journal{\PRL}{86}
{4219}{-4222}{2001}
{Signature change on the brane};
Mars M, Senovilla J M M and Vera R \textit{In preparation}.

\bibitem{commu}Carter B \Journal{\CMP}{17}
{233}{-238}{1970}
{The commutation property of a stationary axisymmetric system};
Carot J, Senovilla J M M and Vera R, \Journal{\CQG}{16}
{3025}{-3034}{1999}
{On the definition of cylindrical symmetry};
Barnes A \Journal{\CQG}{17}
{2605}{-2609}{2000}
{A comment on a paper by Carot et al.}

\bibitem{maseaxconf}Mars M and Senovilla J M M \Journal{\CQG}{10}
{1633}{-1647}{1993}
{Axial symmetry and conformal Killing vectors}

\bibitem{jaumeax}Carot J \Journal{\CQG}{17}
{2675}{-2690}{2000}
{Some developments on axial symmetry}

\bibitem{carter69}Carter B \Journal{\JMP}{10}
{70}{-81}{1969}
{Killing horizons and orthogonally transitive groups in spacetime}

\bibitem{bardeen1970}Bardeen J M \Journal{\AP}{162}
{71}{-95}{1970}
{A variational principle for rotating stars in general relativity}

\bibitem{mps}Vera R \Journal{\CQG}{19}
{5249}{-5264}{2002}
{Symmetry-preserving matchings}

\bibitem{OPSN}Oppenheimer J R and Snyder H \Journal{\PR}{56}
{455}{-459}{1939}
{On continued gravitational contraction}




\end{thebibliography}
\end{document}